\documentclass{PoS}

\usepackage{url}
\usepackage{amsmath,amssymb}
\usepackage{dsfont}

\title{Zeta-regularized vacuum expectation values from quantum computing simulations}

\ShortTitle{Zeta regulated vevs from QC}

\author{\speaker{Karl Jansen}\\
        NIC, DESY Zeuthen, Platanenallee 6, 15738 Zeuthen, Germany\\
        E-mail: \email{karl.jansen@desy.de}}

\author{Tobias Hartung\\
        Department of Mathematics, King's College London, Strand, London WC2R 2LS, United Kingdom\\
        E-mail: \email{tobias.hartung@kcl.ac.uk}}

\abstract{The zeta-regularization allows to establish a
connection between Feynman's path
integral and Fourier integral operator
zeta-functions. This fact can be utilized to perform the
regularization of the vacuum expectation values in quantum field theories.
In this proceeding, we will describe the concept of the
zeta-regularization, give a simple example and demonstrate
that quantum computing can be employed to numerically evaluate
zeta-regulated vacuum expectation values on a quantum computer.}

\FullConference{37th International Symposium on Lattice Field Theory - Lattice2019\\
		16-22 June 2019\\
		Wuhan, China}

\begin{document}

\section{Introduction}

One of the most fundamental concepts in theoretical 
physics is the path integral (or partition function 
in statistical physics) introduced by Feynman
\cite{feynman,feynman-hibbs-styer}
which can provide expectation values

\begin{equation}
\langle \Omega\rangle=\lim_{T\to\infty+i0^+}{\frac{{\rm tr}\left(U(0,T)\Omega\right)}{{\rm tr}\left(U(0,T)\right)}}
\label{pi}
\end{equation}
of important observables $\Omega$ in high energy physics, statistical mechanics and beyond, 
e.g. in turbulence \cite{Margazoglou:2018wqy}.
In eq.~\eqref{pi} $U(0,T)$ is the time evolution operator of a given physical system

\begin{equation}
U(0,T) = \operatorname{Texp}\left(-\frac{i}{\hbar} \int_{0}^{T} H(\tau) d \tau\right)\; ,
\label{timeevo}
\end{equation} 
with $\operatorname{Texp}$ denoting time ordered exponential and $H$ the Hamiltonian 
of the considered physical model. 

A major  
drawback of the path integral as written in 
eq.~\eqref{pi} is, however, that it is ill defined in general. 
A well known way out is to formulate the path integral
on a discrete space-time lattice in Euclidean 
time, see e.g. the textbooks in~\cite{Rothe:1992nt,Gattringer:2010zz}. Although this 
approach of {\em lattice field theory} has been and is still extremely successful in 
computing physical quantities in QCD and other models of high energy physics, 
it is very limited in addressing e.g. 
real time phenomena, questions with a non-zero baryon 
density or CP violation related to matter anti-matter asymmetry of 
the universe. These are very important physical questions where, however, 
a sign problem appears such that standard Markov chain Monte Carlo 
methods are not suitable. 
It would therefore be very important and useful to find a 
regularization
of the path integral in a Lorentzian background metric 
which holds 
non-perturbatively and which leads to a practical way to evaluate  
expectation values in the path integral formalism. 
If such a formalism could be extended to even more general metrics, 
e.g. curved space, then it would lead to an even more powerful exploitation 
of the path integral. 
In this proceeding, 
following refs.~\cite{hartung,Hartung:2018usn,hartung-jansen},   
we present a proposal to provide exactly such a setup.
In particular, there are two main messages to convey: 

\begin{itemize}
\item By a suitable $\zeta$-regularization of the path integral we will 
obtain a mathematically sound, non-perturbative definition of the path
integral which holds in very general metrics and, 
vacuum expectation values computed within this framework are 
{\em physical expectation values}. 
\item For the evaluation of the so defined expectation    
values quantum computations can be used on newly emergent quantum devices.
\end{itemize}

\section{$\zeta$-regularization}

In order to pave the way towards the path integral in the $\zeta$-regularization, 
we remind the reader of the Riemann $\zeta$-function $\zeta_{\rm R}(z)$ which is given by 

\begin{equation}
\zeta_{\rm R}(z) = \sum_{n=1}^\infty \frac{1}{n^z} \; , \{z\in\mathbb{C};\ \Re(z)>1\}\; .
\label{zetaf}
\end{equation}
If we analytically continue $\zeta_{\rm R}(z)$, e.g. through the $\Gamma$-function, 
we receive a mathematically well defined definition of $\zeta_{\rm R}(z)$ for $z\ne 1$. 
In particular, we can choose $z=-1$ to find 

\begin{equation}
\zeta_{\rm R}(-1) = -\frac{1}{12}. 
\label{zetaminusone}
\end{equation}
Although this is clearly a mathematically sound and well controlled result, it is 
rather counter intuitive. At $z=-1$ the series in eq.~\eqref{zetaf} is obviously 
divergent, but nevertheless we obtain a finite number at $z=-1$. This is surprising and we will
come back to this ``mystery'' when we discuss the physical meaning 
of expectation values of observables in the $\zeta$-regularized 
path integral. 

The concept of the Riemann $\zeta$-function can be extended to operators. Let us
look for example at a simple infinite dimensional derivative operator on the torus 
$\mathds{R}/2\pi\mathds{Z}$,

\begin{equation}
{\rm tr} |\partial | \underbrace{=}_{F.T.} \sum_{n=-\infty}^{+\infty} |n| =2 \sum_{n=1}^{\infty} n 
\label{eq:partial}
\end{equation} 
where we performed a Fourier transformation (F.T.) to obtain the sum 
on the right hand side representing the trace of the operator. Again, 
being divergent, 
the series in eq.~\eqref{eq:partial} is ill defined. 

It is, however, possible to define a $\zeta$-trace by introducing a holomorphic family of operators
$\varphi(z)$ 

\begin{equation}
\varphi(z):= |\partial|^{1+z}\;  
\label{eq:zpartial}
\end{equation}
which for ${\rm Re}(z)<-2$ has as a trace

\begin{equation}
{\rm tr} \varphi(z) = 2\sum_{n=1}^{\infty} n^{1+z}\; .
\label{eq:zpartialtrace}
\end{equation}
As in the example of the Riemann $\zeta$-function, the series in 
eq.~\eqref{eq:zpartialtrace}, which represents the trace of the operator 
$\varphi(z)$, can be extended analytically to define the operator 
$\zeta$-function $\zeta(\varphi)(z)$ and ${\rm tr}|\partial|$ can 
be evaluated using the induced {\em $\zeta$-trace}  

\begin{equation}
\zeta(\varphi)(z) = 2\zeta_{\rm R} (-z-1) \Rightarrow {\rm tr} |\partial | := 
\zeta(\varphi)(0) = 2\zeta_{\rm R}(-1) = -\frac{1}{6}\; .
\label{eq:zetatrace}
\end{equation} 
Again, we obtain a well defined value for the $\zeta$-trace of the 
derivative operator but it is certainly not clear at all what 
the physical interpretation of this result would be. 

Let us, in the next step, apply the above ideas to the quantity we are 
interested in, i.e. the time evolution operator 

\begin{equation}
U(0,T) = \operatorname{Texp}\left(-\frac{i}{\hbar} \int_{0}^{T} H(\tau) d \tau\right)\; .
\label{eq:timeevo}
\end{equation}
This time evolution operator has a Fourier integral kernel

\begin{equation}
k(x, y)=\int dr \int d\omega(\xi) e^{i h_{2}(x, y, \xi) r^{2}+i h_{1}(x, y, \xi) r}a(x, y, r, \xi)
\label{eq:fik}
\end{equation}
where we integrate over the spherical coordinates ($d\omega(\xi)$) and the radial 
components ($dr$) separately. In eq.~\eqref{eq:fik} $h_1$ ($h_2$) correspond to 
first (second) order differential operators in the Hamiltonian and the $a(x, y, r, \xi)$ 
represent the ``Fourier 
coefficients''. 

Using the time evolution operator of eq.~\eqref{eq:fik} we can  
{\em formally} define the trace

\begin{equation}
{\rm tr}(U(0,T)) = \int dx \int dr \int d\omega(\xi)
e^{i h_{2}(x, x, \xi) r^{2}+i h_{1}(x, x, \xi) r} a(x, x, r, \xi)
\label{fiktrace}
\end{equation}
which in general is ill defined since the trace can be (and usually is) divergent. 
However, following the spirit of the previous examples, we can introduce 
a family of holomorphic functions $\mathfrak{g}(z)$ obeying  
$\mathfrak{g}(0)=1$ which will eventually be used to define 
the $\zeta$-trace of the Fourier integral kernel. 
Examples of such functions are the choices 
$\mathfrak{g}(z) \propto r^z$. 
But, in principle, $\mathfrak{g}(z)$ can have a much more general form, 
see e.g. section~(2.2).
In mathematical language, introducing $\mathfrak{g}(z)$ means to ``gauge'' the integral kernel 

\begin{equation}
k_{\mathfrak{g}}(x, y)(z)=\int dr \int d\omega(\xi) e^{i h_{2}(x, y, \xi) r^{2}+i h_{1}(x, y, \xi) r}(a\mathfrak{g}(z)(x, y, r, \xi))
\label{fikgauge}
\end{equation}
which allows to define the $\zeta$-trace of the Fourier integral kernel and hence, in turn, 
the time evolution operator 

\begin{eqnarray}
{\rm tr}_\zeta(U(0,T)) & = & \zeta(k_{\mathfrak{g}}(x, y))(z)|_{z=0} \nonumber \\
& = &
\int dx \int dr \int d\omega(\xi) e^{i h_{2}(x, x, \xi) r^{2}+i h_{1}(x, x, \xi) r}(a\mathfrak{g}(z)(x, x, r, \xi))|_{z=0}\; .
\label{eq:fikztrace}
\end{eqnarray}

In a more general setup, as shown in ref.~\cite{hartung}, 
a family of operators $\mathfrak{G}(z)$ with 
the property $\mathfrak{G}(0)=1$ is introduced, 
{\em gauging} the time evolution operator to be of the form 
$U(T,0)\mathfrak{G}(z)$. 
This too leads to the gauged Fourier integral kernel of 
eq.~\eqref{fikgauge} with the corresponding $\zeta$-trace of eq.~\eqref{eq:fikztrace}, 
which in turn allows us to define a $\zeta$-regulated vacuum expectation value

\begin{equation}
\langle\Omega\rangle :=\langle\Omega\rangle_{\zeta} :=
\langle\Omega\rangle_{\mathfrak{G}}(0):=
\lim_{z\to0}\lim_{T\to\infty+i0^+}
\frac{{\rm tr}\left(U(0,T)\mathfrak{G}(z)\Omega\right)}{{\rm tr}\left(U(0,T)\mathfrak{G}(z)\right)}\; .
\label{zetaexpect}
\end{equation}
As proven in \cite{hartung} the expectation value in eq.~\eqref{zetaexpect} 
is now mathematically well defined. However, if we remember the somewhat 
counter intuitive example of the Riemann $\zeta$-function for $z=-1$, 
eq.~\eqref{zetaminusone}, 
the physical meaning 
of the so evaluated expectation value $\langle\Omega\rangle$ in 
eq.~\eqref{zetaexpect} is completely unclear at this point. 

\subsection{Main result}

The main -- and somewhat surprising -- result shown in \cite{Hartung:2018usn}  
is that we actually obtain {\em the physical vacuum expectation value} 
in eq.~\eqref{zetaexpect}. This statement can be summarized in the 
following equations where we denote by $|\psi^{0}\rangle$ the ground state and 
with $|\psi_n^{0}\rangle$ a discretized version of it, see in particular section~3:

\begin{eqnarray}
\langle\psi^{0}|\Omega| \psi^{0}\rangle & = & 
\lim _{n \rightarrow \infty}\left\langle\psi_{n}^{0}, \Omega_{n} \psi_{n}^{0}\right\rangle_{\mathcal{H}} 
\label{main1} \\
& = & \lim_{z \rightarrow 0}\lim _{n \rightarrow \infty} \frac{\left\langle(\mathfrak{G}(z) \Omega)_{n}\right\rangle}{\left\langle\mathfrak{G}(z)_{n}\right\rangle} \label{main2} \\
& = & \lim_{z \rightarrow 0}\lim _{T \rightarrow \infty+i 0^{+}} \frac{\zeta(U(0,T) \mathfrak{G} \Omega)}{\zeta(U(0,T) \mathfrak{G})}(z) \label{main3} \\
& = & \langle\Omega\rangle_{\zeta} \label{main4}\; . 
\end{eqnarray}
Eq.~\eqref{main1} tells us that we obtain the physical expectation value
as the limit $n\rightarrow \infty$ from the expectation 
value taken in a Hilbert space $\mathcal{H}$ evaluated in a suitable discretization scheme. 
Eq.~\eqref{main2} says that this is equivalent to the expectation value 
of the gauged observable taken at $z=0$ in the $n\rightarrow \infty$ limit. 
Eq.~\eqref{main3} states that this 
is in turn equivalent to the $\zeta$-trace of 
the path integral, i.e. the ratio of the gauged observable and time evolution 
operator. Finally, eq.~\eqref{main4} connects this to the 
definition of the $\zeta$-regulated vacuum expectation value. 
Moreover, the main result given above is independent from the 
choice of the gauge assuming the assumptions 
in refs.~\cite{hartung,Hartung:2018usn} are fulfilled. 

\subsection{The free Dirac operator example}

In order to illustrate the steps that lead to the $\zeta$-regulated vacuum 
expectation value let us take the example of the free Dirac operator in 
the continuum 

\begin{equation}
H=\left( \begin{array}{cc}{m c^{2}} & {-i \hbar \sigma_{k} \partial_{k}} \\ {-i \hbar \sigma_{k} \partial_{k}} & {m c^{2}}\end{array}\right) 
\sim \left( \begin{array}{cc}{m c^{2}} & {\hbar r \sigma_{k} \xi_{k}} \\ {\hbar r \sigma_{k} \xi_{k}} & {m c^{2}}\end{array}\right)
\label{dirac}
\end{equation}
where in the second step we have taken the Fourier transform expressed in radial 
($r$) and spherical ($\xi$) coordinates. As a (very simple) gauge, we now consider
the holomorphic function 

\begin{equation}
\mathfrak{g}(z)(x, r, \xi)=r^{z}\; .
\label{diracgauge}
\end{equation}
Other choices of gauges  
gauges are certainly possible. For example, using a positive number~$\delta$, one may choose 
$r^{\delta z}\mathfrak{f}(z)(x,\xi)$, or $(1+r)^{\delta z}\mathfrak{f}(z)(x,\xi)$. 
Furthermore gauges of the form $f(z,r)\mathfrak{f}(z)(x,\xi)$ 
with $f=1$ near $r=0$, $f(z,r)\propto r^{\delta z}$ for $r\gg1$, 
and holomorphic families of continuous functions $\mathfrak{f}$ are common examples. But, it is 
sufficient to consider the simple case of eq.~\eqref{diracgauge} here. 
Inserting eq.~\eqref{dirac} into the general formula of the 
$\zeta$-trace of eq.~\eqref{eq:fikztrace} we obtain 
see, \cite{hartung-jansen}, 

\begin{equation}
\langle H \rangle
=\lim_{z\rightarrow 0}\lim_{T\rightarrow \infty}
\frac{\int d\omega \int dr \left(4mc^2\cos(Tr)-4ir\sin(Tr)\right)r^{z+2}}{\int d\omega\int dr 4\cos(Tr)r^{z+2}}
\label{diractrace}
\end{equation}
which can be evaluated  to

\begin{equation}
\langle H \rangle
=mc^2 - \lim_{z\rightarrow 0}\lim_{T\rightarrow \infty}
\frac{\int dr 4i\sin(Tr)r^{z+3}}{\int dr \cos(Tr)r^{z+2}}\; ,
\label{diracresult}
\end{equation}
with the second term on the right hand side being zero in the limit 
$z\rightarrow 0$ and $T\rightarrow \infty$.
Hence, we obtain for free Dirac fermions $\langle H\rangle = mc^2$. 
Maybe, this appears to be a most complicated way to obtain the 
ground state energy of the free Dirac operator, but this simple 
computation illustrates the steps involved in obtaining 
$\zeta$-regulated vacuum expectation values through solving 
-- possibly very high dimensional -- spherical integrals. More 
examples can be found in \cite{hartung,Hartung:2018usn,hartung-jansen}.

\section{Sketch of proof}

In this section, we want to provide a sketch of the proof of the statement 
that we obtain physical vacuum expectation values. 
A more detailed and mathematically rigorous proof can be found 
in ref.~\cite{Hartung:2018usn} which is based on the results in \cite{hartung}. 
We start with a separable Hilbert space $\mathcal{H}$ and consider the time evolution 
operator 

\begin{equation}
U(t,t') = \mathrm{Texp}\left(-\frac{i}{h} \int_{t}^{t^{\prime}} H(s) d s\right)
\label{inprooftimeevo}
\end{equation} 
where $H$ is the Hamiltonian and $\mathrm{Texp}$ denotes the time ordered exponential. 
Our goal is to compute the vacuum expectation value of an observable $\Omega$ 
in the ground state $\psi^0$ of the Hamiltonian $H$:

\begin{equation}
\langle\psi^{0}|\Omega| \psi^{0}\rangle=\lim _{z \rightarrow 0} \lim _{T \rightarrow \infty+i 0^{+}} 
\frac{\zeta(U(T, 0) \mathfrak{G} \Omega)(z)}{\zeta(U(T, 0) \mathfrak{G})(z)}\; .
\label{proofzetaexpect}
\end{equation}
The proof of eq.~\eqref{proofzetaexpect} proceeds by introducing 
nested, $n$-dimensional subspaces $P_n[\mathcal{H}]$ corresponding 
to orthogonal projections of $\mathcal{H}$ (i.e. measurements). 
These nested sequences are to be taken from a discretization 
scheme which can be a finite or infinite lattice, a complete set of continuous 
functions or more general using a Schauder basis \cite{schauderbasis}. 
The orthogonal projections $P_n$ are to be constructed such that the 
topology of $\mathcal{H}$ is respected (think about topological sectors in QCD).

Next, we construct a Hilbert space $\mathcal{H}_1$
which is densely embedded in $\mathcal{H}$. The subspaces  
$P_n[\mathcal{H}]$ are to be contained 
in $\mathcal{H}_1$ and the gauge $\mathfrak{G}(z)$ as a well as the gauged observable
operator $\mathfrak{G}(z) \Omega$ are to be bounded operators from 
$\mathcal{H}_1$ to $\mathcal{H}$ provided that ${\rm Re}(z)$ is less than 
some positive real number. 

If we consider a Hilbert space $\mathcal{H}$ as a $L_2$-space, then 
$\mathcal{H}_1$ will in general be a Sobolev space $W_2^s$, i.e. 
a space of functions that admit derivatives up to order $s$ 
--where $s$ is to be taken sufficiently large-- and 
have an (integral) norm of order $2$, i.e. in one dimension
$\|f\|_{k,2}=\left(\sum_{i=0}^{k} \int\left|f^{(i)}(t)\right|^{2} d t\right)^{\frac{1}{2}}$.
Sobolov spaces are commonly used in the field of differential equations and we may 
think of taking the differential operator there (e.g. the 
Laplace operator) to play the role of the Hamiltonian in our 
problem. 

Having constructed $\mathcal{H}_1$, we need the orthogonal 
projections $Q_n$ onto $P_n[\mathcal{H}]$ in $\mathcal{H}_1$. This finally allows
us to discretize the observable $\Omega$ of interest

\begin{equation}
\Omega_n := P_n \Omega Q_n
\label{discreteA}
\end{equation}
and correspondingly the 
discretized time evolution operator 

\begin{equation}
U_n(t,t')=\operatorname{Texp}\left(-\frac{i}{\hbar} \int_{t}^{t^{\prime}} P_{n} H(s) Q_{n} d s\right)
\label{discreteU}
\end{equation}
which is then in $P_n[\mathcal{H}]$. 
Note that by ``discretization'' we mean here a very general discretization 
scheme which can be a finite or infinite lattice, a function basis (think of 
Fourier modes) or elements of a Schauder basis which can can be considered 
``generalized'' Fourier modes once the basis is orthonormalized.  

The discretized ground state wave function is obtained by the minimum of the energy, i.e. 

\begin{equation}
\psi_n^0 = \underset{\left\lVert\Psi_n\right\rVert=1}{\rm argmin} \langle \Psi_n | H_n | \Psi_n\rangle
\label{discretepsi}
\end{equation}
where $H_n = P_n H Q_n$ living in $P_n[\mathcal{H}]$. 
It is then possible to show \cite{hartung,Hartung:2018usn} that the discretized 
ground state $\psi^0_n$ converges to the true ground state $\psi^0$.

\begin{equation}
\left\langle\psi_{n}^{0}, \psi^{0}\right\rangle_{\mathcal{H}} \psi_{n}^{0} 
\rightarrow \psi^{0}\quad(n\rightarrow\infty) .
\label{groundstate}
\end{equation}
In other words, 
eq.~\eqref{groundstate} states, fairly untrivially, 
that the discretized theory reproduces the infinite dimensional  
continuum theory in the limit $n\rightarrow \infty$ for a suitably chosen discretization scheme.

In order for this statement to hold, we need the underlying quantum field theory to 
have an energy gap, that $H$ is self-adjoint, and that $H$ plays the role of a generator of 
the exponential function in the time evolution operator. 
In a more general language, the one parameter (here the time~$t$) exponential of an operator 
is considered as a semi-group and the time-dependent Hille-Yosida theorem (see theorem 5.3.1
in \cite{pazy}) 
states 
under which condition the operator valued exponential has a well defined 
meaning, which is based on the spectrum of the considered operator.

In the discretized setup the expectation value of an observable $\Omega_n$ is given 
by 
\begin{equation}
\left(\psi_{n}^{0}\left|\Omega_{n}\right| \psi_{n}^{0}\right\rangle=
\lim _{T \rightarrow \infty+i 0^{+}} 
\frac{\operatorname{tr}\left(U_{n}(T, 0) \Omega_{n}\right)}{\operatorname{tr}\left(U_{n}(T, 0)\right)}
\label{discreteAn}
\end{equation}
which takes over to the discretized gauged observable 

\begin{equation}
\frac{\left\langle\psi_{n}^{0}\left|(\mathfrak{G}(z) \Omega)_{n}\right| \psi_{n}^{0}\right\rangle}{\left\langle\psi_{n}^{0}\left|\mathfrak{G}(z)_{n}\right| \psi_{n}^{0}\right\rangle}
=\lim _{T \rightarrow \infty+i 0^{+}} \frac{\operatorname{tr}\left(U_{n}(T, 0)(\mathfrak{G}(z) \Omega)_{n}\right)}{\operatorname{tr}\left(U_{n}(T, 0) \mathfrak{G}(z)_{n}\right)}\; .
\label{discreteAgauged}
\end{equation}
Now, through the gauging procedure the operators 
$U(T, 0) \mathfrak{G}(z)\Omega$ 
and 
$U(T, 0) \mathfrak{G}(z)$  
are of trace class in $\mathcal{H}$ for ${\rm Re}(z) \ll 0$. 
Furthermore, through the construction of $\mathcal{H}_1$ 
and $H$ satisfying the Hille-Yosida theorem we
can deduce that we can take the limits

\begin{eqnarray}
\operatorname{tr}\left(U_{n}(T, 0)(\mathfrak{G}(z) \Omega)_{n}\right) & \rightarrow & 
\operatorname{tr}(U(T, 0) \mathfrak{G}(z) \Omega )\nonumber \\
\operatorname{tr}\left(U_{n}(T, 0) \mathfrak{G}(z)_{n}\right) & \rightarrow & \operatorname{tr}(U(T, 0) \mathfrak{G}(z))\; .
\label{limit1}
\end{eqnarray}
These limits allow us now for ${\rm Re}(z) \ll 0$ to identify the 
$\zeta$-regularized vacuum expectation values 

\begin{equation}
\frac{\left\langle\psi_{n}^{0}\left|(\mathfrak{G}(z) \Omega)_{n}\right| \psi_{n}^{0}\right\rangle}{\left\langle\psi_{n}^{0}\left|\mathfrak{G}(z)_{n}\right| \psi_{n}^{0}\right\rangle} 
\rightarrow \lim _{T \rightarrow \infty+i 0^{+}} \frac{\zeta(U(T, 0) \mathfrak{G} \Omega)(z)}{\zeta(U(T, 0) \mathfrak{G})(z)}\; .
\label{expect1}
\end{equation}

In the steps above we needed the condition that ${\rm Re}(z) \ll 0$ because \eqref{limit1} only holds for 
trace-class operators. 
As a last step in the proof we need to extend 
eq.~\eqref{expect1} to all values of $z$ with ${\rm Re}(z)<\epsilon$ 
when $\epsilon>0$ is chosen\footnote{This is strictly speaking more restrictive than 
necessary although it is very difficult to construct an example which would not satisfy this.}.
For this to be true, we need that $\mathfrak{G}$ and $\mathfrak{G}\Omega$ 
are bounded operators for ${\rm Re}(z)$ bounded by the same positive 
real number $\epsilon$ again. A second requirement is that the norms
$\|\mathfrak{G}(z)\Omega\Psi_n^0\|_{n\in \mathbb{N}}$  
and 
$\|\mathfrak{G}(z)\Psi_n^0\|_{n\in \mathbb{N}}$  
are bounded for ${\rm Re}(z)<\epsilon$ with the same $\epsilon$ as above. 
Under these conditions, the limit in eq.~\eqref{expect1} holds 
for all values of $z$ in a subset $\mathcal{D}\subseteq\{z \in\mathbb{C}; {\rm Re}(z) < \epsilon\}$  
such that the set $\{z\in\mathbb{C};\ {\rm Re}(z)<\epsilon \text{ and }z\notin\mathcal{D}\}$ 
contains only isolated points which guarantees the uniqueness of the analytical continuation.
In more general terms, the convergence in eq.~\eqref{expect1} follows 
from the so-called Vitali Porter theorem\footnote{According to 
ref.~\cite{schiff} this theorem was proven independently by Vitaly \cite{vitali} 
(also published in \cite{vitali2}) and Porter~\cite{porter}.} 
see \cite{schiff}  chapter 2.4
which is a convergence 
theorem, well known in the mathematical literature which, applied to our problem at hand, 
guarantees the convergence in eq.~\eqref{expect1}.  
In particular, as we show in \cite{Hartung:2018usn}, the here 
proposed $\zeta$-regularized vacuum expectation value fulfills the 
conditions of the Vitali Porter theorem and hence the convergence
in eq.~\eqref{expect1} holds.

A subtlety in the proof is, whether $z=0$ is indeed in the subset $\mathcal{D}$. However, 
in \cite{Hartung:2018usn} it was proven that this is indeed the case.
As stated in the main result, the fact that we have the convergence 
of eq.~(\ref{main1}) we finally can relate the $\zeta$-trace of the 
gauged, $\zeta$-regulated path integral to the physical 
expectation value. 

\subsection{A note on the anomaly}

The claim of our result in eq.~(\ref{main3}) is that we can use the path integral in 
continuous Minkowski 
space time and obtain finite, well defined expectation values that correspond to 
the desired physical vacuum expectation values. 
This leads to the interesting question
how the axial anomaly appears in this setup and how chiral gauge 
theories can be constructed within this approach. The short answer is 
that, if gauges are used that respect the (chiral) symmetry of the 
original theory, then the anomaly will appear through the 
fact that the measure is not invariant. In other words, the derivation
of the anomaly will follow closely the construction of Fujikawa 
in \cite{Fujikawa:1986hk}. To make this statement more solid and put it on 
a rigorous mathematical ground is subject to a work we plan to carry 
out in the future.

\section{Quantum Computing $\zeta$-regulated vacuum expectation values}

The theoretical notion of a mathematically well defined 
vacuum expectation value in the path integral or in a suitably chosen 
discretization scheme leaves open 
the question, how the so obtained vacuum expectation values can 
be computed in practice. One direction would be to develop highly 
efficient methods for solving high dimensional 
integrals in spherical coordinates, along the line of 
refs.~\cite{Nied92,KSS_Review12,Jansen:2013jpa,Ammon:2015mra,Ammon:2016jap}. 
Another, very appealing and intriguing
possibility is to employ quantum computations for this purpose. 

In the last years, quantum computers have emerged as 
promising novel technologies that could be used to solve 
problems that are extremely difficult or even impossible 
to address classically. The usage of quantum computers 
has changed in the last years and superconducting qubit architectures such 
as the ones at Rigetti \cite{rigetti}, IBM \cite{ibmq}, or
D-Wave \cite{dwave} are now remotely accessible 
devices which can be programmed in Python with software libraries 
supplied by the companies. 
Example programs and instructions can be found on the corresponding 
web-pages. Also simulators are provided which can be used 
locally. This makes it easy to develop and test programs for correctness. The simulators also have the option to  
switch on noise which provides good estimates  
of the performance on the real hardware. 

Such a setup allows to use quantum computers
``easily'' nowadays and provides the opportunity to test the
potential of quantum computing for the particular problem
one is interested in. 
For our case of $\zeta$-regularized vacuum expectation values 
we will employ the method of variational quantum simulation, 
as explained in more detail below, for the problem of the 
1-dimensional hydrogen atom. 
For the calculation we will actually employ  
eq.~\eqref{main1} of the main result stated above. This means that 
we will use a suitable discretization scheme to quantum compute the expectation 
value of the Hamiltonian of the 1-dimensional hydrogen atom 
in the ground state.

\subsection{One dimensional hydrogen atom} 

As anticipated above, as an example for a quantum computation
we will consider here the 1-dimensional hydrogen atom
with the Hamiltonian

\begin{equation}
H=-\frac{\partial^{2}}{2 m}+ qU(x)\;,\;\;
U(x)=\left\{\begin{array}{ll}{x} & { ; x \in(0, \pi)} \\ {0} & { ; x \in(-\pi, 0]}\end{array}\right.
\label{hydrogen1}
\end{equation}
where we restrict ourselves to the interval $(-\pi, \pi)$. 
The Hilbert space is $L_2(-\pi,\pi)$ and the 
orthogonal projection $P_n$ maps onto 
the finite dimensional linear space 

\begin{equation}
{\rm lin}\left\{\varphi_k;\ -\left\lfloor\frac{n}{2}\right\rfloor\le k\le n-1-\left\lfloor
\frac{n}{2}\right\rfloor\right\}\ (\text{so that }\dim P_n[\mathcal{H}]=n)
\label{hydrogen2}
\end{equation}
and where $\varphi_k(x)=\frac{1}{\sqrt{2\pi}}e^{ikx}$.

\begin{figure}[t]
\centering
\includegraphics[width=7cm]{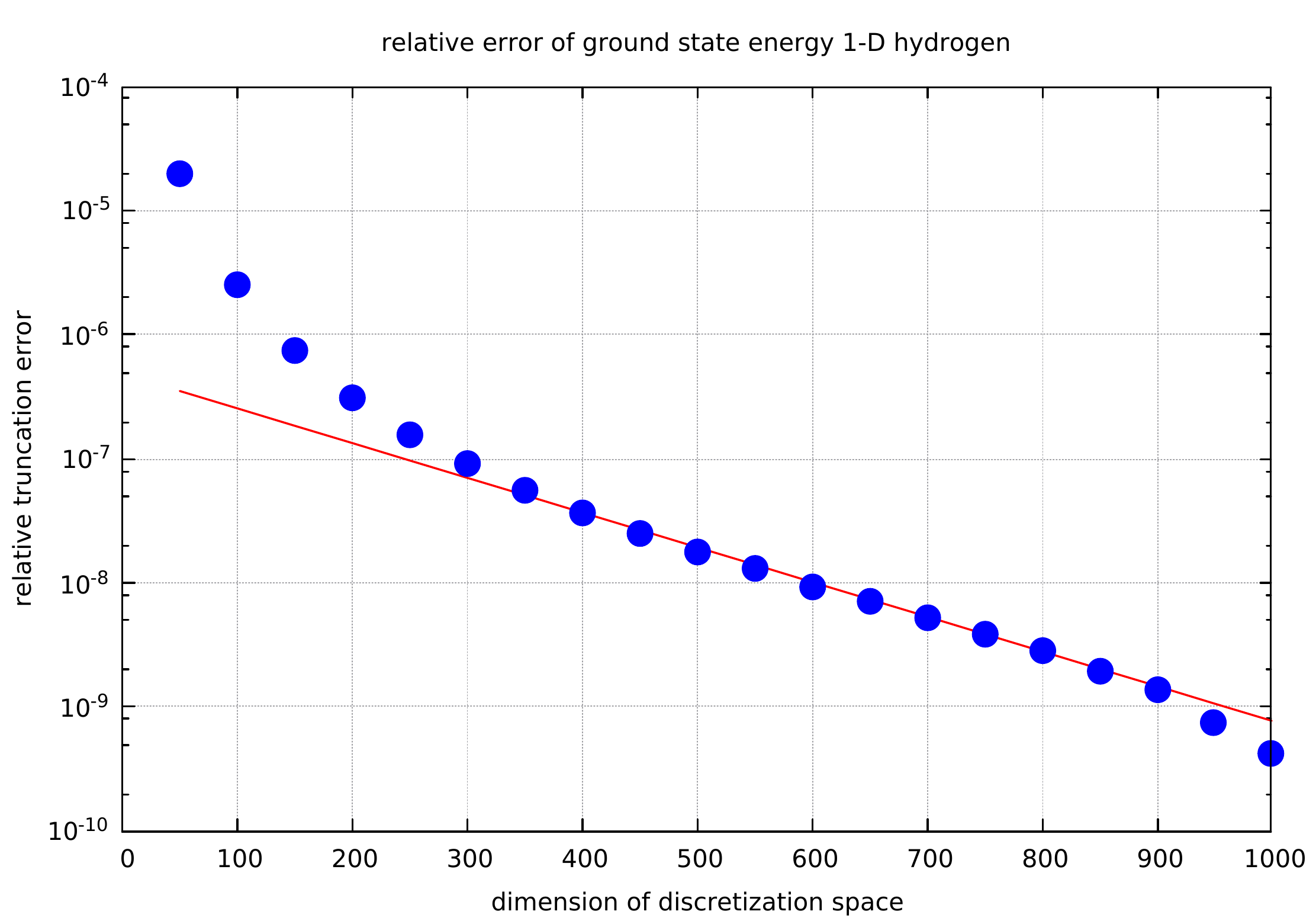}
\caption{Results from exact diagonalization. We show 
the relative error $\Delta$ of the energy $E$ when using a dimension $n$ 
compared to the the largest dimension used, i.e. $n=1024$, 
i.e. $\Delta = \frac{E(n)-E(1024)}{E(1024)}$.}
\label{fig:exactdiag}
\end{figure}

The matrix elements of
the 1-dimensional Hamiltonian of the hydrogen atom can be computed analytically and we obtain
for $k=l$

\begin{equation}
  \left\langle\varphi_{k}, H \varphi_{k}\right\rangle =\frac{k^2}{2m}+\frac{q\pi}{4}
\end{equation}
and for $k\ne l$
\begin{equation}
\left\langle\varphi_{l}, H \varphi_{k}\right\rangle =\frac{q\left((-1)^{k-l}(1-i \pi(k-l))-1\right)}{2 \pi(k-l)^{2}}\; .
\label{hydrogen3}
\end{equation}
Having these matrix elements at hand, we can construct the full matrix which 
can be exactly diagonalized,  
if the dimension $n$ is not too large. In fig.~\ref{fig:exactdiag} we show 
the relative error of the lowest eigenvalue, i.e. the ground state energy, with respect to 
the largest dimension $n=1024$ we have used. The nice result of fig.~\ref{fig:exactdiag} is that 
we see an exponentially fast convergence to the $n=\infty$ ground state energy. 
This is a promising result since it points to the possibility that in 
practice only a small number of qubits are required to obtain the 
ground state energy and wave function to a good accuracy.

In order to write the Hamiltonian of the hydrogen atom in a form 
that can be used on a quantum computer, we need to write  
it in terms of the Pauli matrices
$\sigma^1=\sigma_x,\ \sigma^2=\sigma_y,\ \sigma^3=\sigma_z,$ and $\sigma^0=\mathds{1}_{2\otimes 2}$
such that the set 
$\left[ \sigma^0, \sigma^1, \sigma^2, \sigma^3\right]$
span 
a {\em Pauli basis}
\begin{equation}
\left\{S^{q}=\sigma^{q_{{Q}-1}} \otimes \sigma^{q_{Q-2}} \otimes \ldots \otimes \sigma^{q_{0}} ; q \in 4^{Q}\right\}
\label{paulibasis2}
\end{equation}
with the exponent of four appearing because we have, in general, the possibility of
all four matrices from the set $\left[ \sigma^0, \sigma^1, \sigma^2, \sigma^3\right]$.

Having the matrix elements of eq.~\eqref{hydrogen3} at our disposal 
we can now project them onto the Pauli basis, and obtain the qubit 
Hamiltonian $H_{Q}$. Note that $H_Q$ corresponds to $H_n$ with $n=2^Q$ in eq.~\eqref{discreteA}

\begin{equation}
H_{Q}=\sum_{q \in 4^{Q}} \frac{{\mathrm tr}\left(H_{2^Q} S^{q}\right)}{2^{Q}} S^{q}\; .
\label{paulibasis3}
\end{equation}
$H_{Q}$
is now suited to be implemented on a quantum computer. 
The steps described above and the qubit 
Hamiltonian in eq.~\eqref{paulibasis3} are very general and can be used 
when the matrix elements 
$\left\langle\varphi_{l}, H \varphi_{k}\right\rangle$ can be evaluated  
for a given Hamiltonian $H$. 
However, it needs to be stressed that in simpler cases, e.g. when 
employing the Ising or the Heisenberg model, the qubit 
Hamiltonian can be constructed in a much more direct and simpler way. 

\subsection{Variational quantum simulation} 

One frequently used way to obtain the ground state energy and wave function 
on a quantum computer is the method of variational quantum simulations 
see, e.g.~\cite{peruzzo}.
In this approach, an initial state vector $|\Psi_{\rm init} \rangle$ is generated first. 
On this state vector a sequence of gate operations is applied which are 
single qubit unitary operations $e^{-iS\theta}$ depending on a parameter 
$\theta$ and with $S$ being the Pauli matrices or, e.g., a Hadamard 
gate. Moreover, we can have entanglement gates such as a (parametric) CNOT gate. 

In this way, a state vector $|\Psi(\vec{\theta})\rangle$ depending on all 
parameters of the applied unitary gate operations is generated, 

\begin{equation}
|\Psi(\vec{\theta})\rangle=e^{-i S_{(n)}\theta_{n}} \dots e^{-i S_{(1)}\theta_{1}} |\psi_{\mathrm{init}}\rangle\; .
\label{var1}
\end{equation}
Defining $R_{j} :=e^{-i S_{(j)}\theta_j}$  
a (energy) cost function $C$ can be computed 

\begin{equation}
C:=\left\langle\psi_{\mathrm{init}}\left|\left(\prod_{j=1}^{n} R_{j}\right)^\dagger H \prod_{j=1}^{n} R_{j}\right| \psi_{\mathrm{init}}\right\rangle\; .
\label{var2}
\end{equation} 
The goal is then to minimize 
this cost function over the vector of parameters $\vec{\theta}$. 
The employed strategy is to evaluate the cost function on the quantum 
computing hardware while the minimization over the variational 
parameters is performed on a classical computer.

In our original work \cite{Hartung:2018usn}, we tried to use the libraries that were 
provided in the python software package of Rigetti to minimize
the cost function built form the 1-dimensional hydrogen Hamiltonian,  
see refs.~\cite{rubin,rigetti2} for a discussion of variational
eigensolvers within the Rigetti framework. 
However, in todays NISQ (Noisy Intermediate Scale Quantum computers) area, 
we were not successful to obtain any meaningful result on the real hardware. 

We therefore resorted to a more straightforward method consisting of a 
sequential update of single parameters $\theta_k$.
In particular, we divided each $\theta_k \in [0,2\pi]$ into $N$ steps 

\begin{equation}
\theta_k = 2\pi i/N \pm \epsilon\; , \; k=1,...,N
\label{var3}
\end{equation}
with $\epsilon < \pi/N$ some random noise. We then visited each $\theta_k$ separately, 
minimized the energy for this $\theta_k$ and proceeded to the next angle. 
Sweeping in this way several times over all angles $\theta_k$ allowed us  
to eventually obtain the targeted minimum of the cost function in eq.~\eqref{var2}. 

For the quantum hardware computations we used Rigetti's 8 qubit Agave chip. 
In particular, we selected the qubit with the best measured fidelity 
and implemented 
the qubit Hamiltonian of eq.~\eqref{paulibasis3}. 
While this approach worked for 1 qubit successfully and we could obtain 
an about 95\% accuracy for the ground state energy, we were not able to achieve 
a significant result for 2 qubits.

In order to improve this --rather disappointing-- result
we implemented a gradient descent algorithm to find the minimum of the 
cost function in eq.~\eqref{var2}.
The 
gradient of the cost function can be obtained through the differentiation 
with respect to the parameters $\theta_k$ and the gradient vector can then actually 
be computed on the quantum hardware itself. 

Let us for illustration purposes consider the cost function for 2 unitaries only

\begin{equation}
C:=\left\langle\psi_{\mathrm{init}}\left|\left(e^{i\sigma^1\theta_1}e^{i\sigma^2\theta_2}\right)^\dagger H e^{i\sigma^1\theta_1}e^{i\sigma^2\theta_2}\right| \psi_{\mathrm{init}}\right\rangle\ .
\label{gradeint1}
\end{equation} 
Then the gradient vector is obtained as

\begin{equation}
D= \left(\frac{\partial C}{\partial \theta_1}, \frac{\partial C}{\partial \theta_2}\right)\; .
\label{gradient2}
\end{equation} 
E.g., the derivative with respect to $\theta_1$ is given by

\begin{equation}
\frac{\partial}{\partial\theta_1} C=\left\langle \psi_{\mathrm{init}}\left|\left(e^{i\sigma^1\theta_1}e^{i\sigma^2\theta_2}\right)^\dagger
\left[ (i\sigma^1)^\dagger H + H(i\sigma^1) \right] e^{i\sigma^1\theta_1}e^{i\sigma^2\theta_2} \right| \psi_{\mathrm{init}}\right\rangle\; ,
\label{gradeint3}
\end{equation} 
where we commuted the Pauli matrix $\sigma^1$ through the unitaries, when necessary.  
This means that we can re-use the already generated state vector and measure 
$(i\sigma^1)^\dagger H + H(i\sigma^1)$ directly on the quantum hardware. 
A gradient descent algorithm can then be constructed 
to obtain a new state vector $\vec{\theta}^{{\mathrm new }}$ through 

\begin{equation}
\vec{\theta}^{{\mathrm new }} :=\vec{\theta}^{{\mathrm old }}-\eta \nabla C\left(\vec{\theta}^{{\mathrm old }}\right)
\label{gradient4}
\end{equation} 
with a tunable learning rate $\eta$. Note that the steps above can be generalized to 
much more complicated circuits. For example, when a CNOT gate is used, it can be
expressed in terms of Pauli matrices again.  
Hence, the Pauli matrix originating from a differentiation can again be commuted through 
the CNOT entanglement gate and we find a similar structure as in eq.~\eqref{gradeint3}, although 
with a more complicated operator to be measured for the evaluation of the gradient. 

With the approach of this {\em global} gradient descent algorithm and Rigetti's 
new hardware \cite{rigetti} we could get results also for 2 qubits
with a 90\% fidelity for the ground state wave function. However, the 3-qubit case 
was unfortunately not successful. 
In addition, the global algorithm sketched above is suitable for a small number of 
qubits only. For many qubits, the measurements 
become very difficult since the scaling of the algorithm is exponential. Hence, 
alternative new methods and algorithms need to be developed and work in this direction is in 
progress.

\section{Conclusion} 

In this proceeding we have shown that the $\zeta$-regularization leads  
to 
a mathematically well defined, non-perturbative expression 
of Feynman's path integral in a Minkowski (or more general) 
metric. 
Within this $\zeta$-regularization vacuum expectation 
values can be computed and we sketched a proof that these are the 
physical vacuum expectation values. 

From a practical side, a 
vacuum expectation value can be computed in two ways based 
on our main result, eqs.~\eqref{main1}-\eqref{main4}. 
When using eq.~\eqref{main3} in the Fourier integral kernel representation 
a high dimensional spherical integral
needs to be solved. 
When employing eq.~\eqref{main1}   
the vacuum expectation value is obtained through a limit procedure of 
an appropriate discretization scheme. Here, the ground state energy and 
wave function can be computed   
--at least in principle-- either 
by tensor network methods or by employing a quantum computer, 
see ref.~\cite{Banuls:2019rao} and ref.~\cite{Banuls:2019bmf} for applications 
in high energy physics, respectively. 

In the present work we evaluated the ground state energy of a 1-dimensional 
hydrogen atom on a quantum computer through the hybrid classical-quantum approach of a variational 
quantum eigensolver. Although we could demonstrate that 
this method works in practice to compute the energy as a vacuum 
expectation value, the number of qubits that we could use has been very small. 

The here presented framework of the $\zeta$-regularization 
of the path integral is still novel and only very first steps 
have been taken to explore this approach to quantum field theories. 
In order to develop both, the conceptual as well as the practical 
aspects of the $\zeta$-regularization further, 
a number of main steps (our ``wishlist'') 
need to be developed in the future:

\begin{itemize}
\item It would be very desirable to explore more problems where the 
$\zeta$-regularization can be exploited from a theoretical point of view. Examples
are the application in perturbation theory, curved space-time, 
or looking at chiral gauge theories. 
\item From a practical side, in order to evaluate the 
$\zeta$-regulated path integral, efficient methods are needed to solve high-dimensional spherical integrals 
to obtain expectation values of physical observables in a quantitative way. 
\item Working in a Hamiltonian approach using a suitable 
discretization scheme, novel algorithms for tensor networks and 
quantum computations need to be developed. In particular, in the latter 
case algorithms are needed which are 
efficient, scalable, and robust against errors of the used quantum device.   
\item Finally, better quantum hardware need to be built with a high qubit fidelity and 
optimally even with error correction. 
\end{itemize} 

\section*{Acknowledgment}
We thank P. Stornati for very useful discussions on the gradient descent 
part of this work.


\bibliographystyle{unsrt}
\bibliography{bibliography}
%

\end{document}